\documentclass[aps,prb,twocolumn,groupedaddress,floatfix,showpacs]{revtex4-2}

\usepackage{graphicx}
\usepackage{amsmath,amssymb}
\usepackage{subcaption}
\usepackage{color}
\definecolor{green}{rgb}{0,0.5,0}
\begin{document}


\title{Propagation processing of short pulses in Rydberg exciton medium\\ under blockade conditions}


\author{Sylwia Zieli\'{n}ska-Raczy\'{n}ska}
\author{David Ziemkiewicz}
\email{david.ziemkiewicz@utp.edu.pl}
 \affiliation{Department of
 Physics, Technical University of Bydgoszcz,
\\ Al. Prof. S. Kaliskiego 7, 85-789 Bydgoszcz, Poland}


\date{\today}




\begin{abstract}
Propagation of short pulses through Cu$_2$O crystal containing Rydberg excitons is studied with the use of density matrix formalism and FDTD method. Saturation effects related to the so-called Rydberg blockade are studied extensively, exploring not only reduction of absorption (bleaching) but also power-dependent changes of the dispersive properties of the medium. The role of exciton lifetime and coherent population oscillations in the dynamics of the system is investigated. A pump-probe setup with two pulses is also studied, showing good agreement with recent experimental studies.
\end{abstract}



\maketitle

\section{Introduction}
Coulomb-bound pairs of one electron and one hole called excitons, are often seen as the low-energy counterpart of the hydrogen atom. This picture has been successfully invoked to describe photoexcitation of a semiconductor in low-density limit. Nevertheless, there are significant limitation to the atomiclike description of excitons. These quasiparticles are complex many-body states embedded in a background of a crystal lattice, which interact via scattering and screening. Excitons generally exist in a mixture of unbound charge carriers. Rydberg excitons (RE) first observed in Cu$_2$O by Kazimierczuk \emph{et al} \cite{Kazimierczuk}, unlike those  low-lying excited states, exhibit unique properties including huge dimensions reaching micrometers for excitons with principal quantum numbers $n>20$, strong dipole-dipole interactions, long radiative lifetimes of hundreds nanoseconds. All RE peculiar parameters scale strongly with  powers of the principal quantum number $n$, i.e. radius $r \sim n^2$, lifetime $\tau \sim n^{3}$ \cite{scaling}.
The resulting dipole moments lead to giant exciton interactions, which manifest in a interaction-induced nonlinear response of Rydberg excitons observed already even at low light intensities \cite{myKerr, PRL22}. Such  a long-range mutual interaction between individual Rydberg excitons $ V_{vdW} \sim n^{11}$ is a source of shifting the resonance energy of a one of them out of the excitation linewidth, which results in a phenomenon called Rydberg blockade. This prevents the excitation of other excitons in the vicinity, within a certain volume  $V_{bloc} \sim n^{7}$ and results in a  reduction of absorption at the excitonic resonance. Rydberg blockade can establish strongly correlated states and is a key concept in quantum sensing \cite{Heckotter2024} and single-photon sources \cite{Khazali}.
Rydberg states of excitons are promising quantum objects to engineer controllable optical devices in a solid-state system. For this purpose, a deeper understanding of the dynamics of Rydberg excitons and of their potential for coherent manipulation by light becomes important.
 Rydberg blokade is also an important benchmark in implementation of Rydberg excitons  in quantum information processing with potential applications for realizing quantum logic gates. However the fundamental stage to realize these tasks is  a thorough investigation of dynamical problems of light pulses  propagation in the Rydberg excitons media and to examine the influence of Rydberg blockade on coherent processing of laser pulses in RE media.

So far, most of the works on Rydberg excitons physics have been devoted to investigate experimentally and theoretically their spectroscopic  characteristic. The high-resolution absorption, photoluminescence and nonlinear spectroscopy  reveal the remarkable subtle exciton features \cite{Assmann2020, Naka17}, but recently scientific interest has been  focused on same  dynamical aspects of systems with REs \cite{Panda,Orfanakis22,Thomas,myQB,Thomas2}.

The dynamics of REs system in a ladder configuration with time-resolved two-color pump-probe spectroscopy was examined both theoretically \cite{Rotteger} and experimentally \cite{Heck2023,Panda}, where the temporal evolution of RE systems  and a dependence on pump pulse intensity were discussed and complexity of the decay processes was analyzed. 

On the other hand, the studies of Rydberg blockade are limited to a steady-state conditions. Morawetz \emph{et al} \cite{Morawetz2023} calculated the probe-pump absorption spectra to stimulate the main outcome of experiments discussing the influence on interexcitonic interaction. The scaling dependencies presented in \cite{Morawetz2023} will serve as a groundwork for the results presented in this paper, extending the theoretical treatment of Rydberg blockade beyond steady-state studies and continuous wave illumination by a monochromatic laser field.

However, very recently Minarik \emph{et al} \cite{Minarik} also performed time-resolved femtosecond probe-pump experiment in Cu$_2$O; they observed spectral signature of Rydberg blockade in shaping Rydberg exciton dynamics.

In this paper, we focus on dynamics of the RE system when illuminated by a short near-resonant pulse or a sequence of pulses. Depending on the light intensity, the Rydberg blockade for a given state might have an essential and dynamical influence on excitation population. Our theoretical study, which combines Monte Carlo approach presented in \cite{Morawetz2023} with time-dependent evolution of density matrix  will provide insight into dynamics of optical pulse propagation through Rydberg exciton medium indicating the role of RE lifetimes, power dependent saturation effects and Rydberg blockade influence on propagation characteristic.
   
\section{The general theory}
In this section we outline a model of our system, which includes  the minimum possible complexity to illustrate the physical principle of the theoretical approach, i.e. the time evolution of the density matrix. The inclusion of Rydberg blockade which affects the action for higher excitonic states and higher light intensities resulting in changes of propagation conditions and therefore turns out to be essential for the process.

\subsection{Propagation of laser pulses in Rydberg excitons media}

\begin{figure}[ht!]
\includegraphics[width=.5\linewidth]{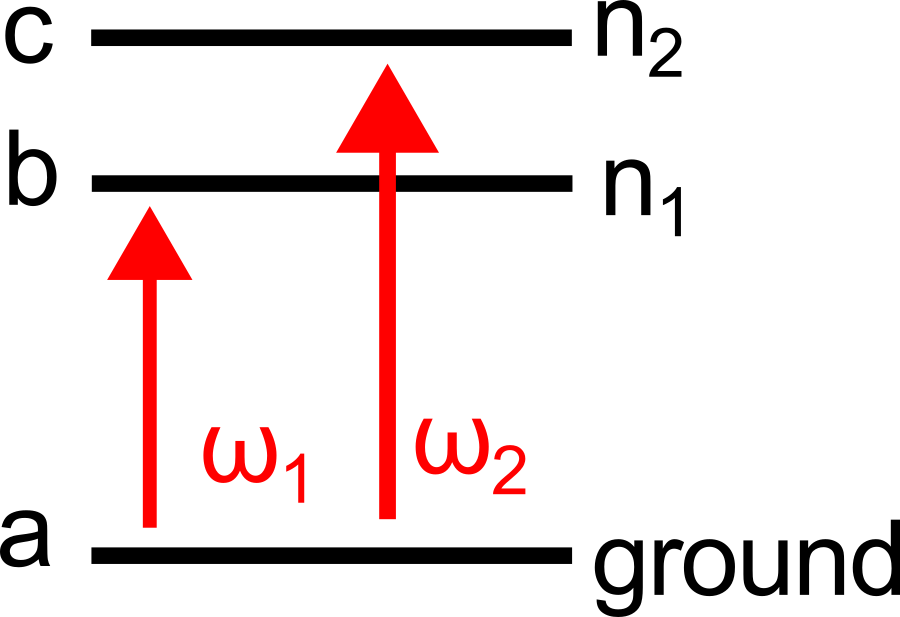}
\caption{Energy level schematic of the considered system}\label{fig:drabinka}
\end{figure}

We consider a quasi-one dimensional Cu$_2$O crystal irradiated by one or two laser pulses presented on Fig. \ref{fig:drabinka}. The medium has
$V$ configuration with ground state $a$ being coupled with two excited states $b$ and $c$ by laser pulses 1 and 2, respectively.

A single exciton resonance is modeled by two states: the excited state $|j\rangle$  and the excitonic vacuum as a ground state $|a\rangle$. Both states are coupled by the laser
 pulse of a frequency $\omega_{i}$, a wave vector $k_i$, $i$=1,2, with amplitudes $\varepsilon_i$ and polarization $\textbf{e}$
\begin{equation}\label{eq:field}
\textbf{E}_i(z, t)= \textbf{e}\varepsilon_i(z,t)\exp[i(\omega_{i} t-k_i z )].
\end{equation} 
The time evolution of the excitonic density matrix $\rho=\rho(z,t)$ for a single exciton at position $z$ is governed by von Neumann equation  with a phenomenological dissipative term R \cite{kossak,myQB} 
\begin{equation}\label{eq:vN}
i\hbar \dot{\rho}=[H,\rho]+R\rho.
\end{equation}
The Hamiltonian of such a system interacting with an electromagnetic field reads
\begin{eqnarray}
&&H=H_0+V+H_{vdW}=E_a|a\rangle \langle a|+\sum_{j=b,c} E_j|j \rangle \langle j|\nonumber\\
&&+\sum_{j=b,c} 2\Omega_j\hbar cos(\omega t)[|a\rangle \langle j|+|j \rangle \langle a|]+H_{vdW},
\end{eqnarray}
where 
\begin{equation}\label{eq:omegaj}
\Omega_j=\frac{\varepsilon d_{aj}}{2\hbar}
\end{equation}
is the Rabi frequency for the $a \rightarrow j$ transition, $d_{aj}$ are transition dipole moments and $E_{j}=\hbar \omega_{1,2}$ are energies of exciton states. 
 The influence of Rydberg blockade results in strongly correlated many-body states of Rydberg excitations \cite{nature2021}, and is described by the van der Waals potential in the form
\begin{equation}\label{eq:HvdW}
H_{vdW}=\sum\limits_{\mu}\frac{C^\mu_6}{R^6_{ij}}|\mu \rangle\langle\mu|,
\end{equation}
where $|\mu\rangle$ is a two-exciton eigenstate \cite{Walther2018}.

The set of equations describing a time evolution in V-configuration has the following form
\begin{eqnarray}
i\dot{\rho}_{ab}&=&(E_a-E_b)\rho_{ab}+V_{ab}\rho_{bb}+V_{ac}\rho_{cb}-V_{ab}\rho_{ab}\nonumber\\
i\dot{\rho}_{ac}&=&(E_a-E_c)\rho_{ac}+V_{ab}\rho_{bc}+V_{ac}\rho_{cc}-V_{ac}\rho_{aa}\nonumber\\
i\dot{\rho}_{bc}&=&(E_b-E_c)\rho_{bc}+V_{bc}\rho_{ac}-V_{ac}\rho_{ba}\nonumber\\
i\dot{\rho}_{aa}&=&V_{ab}\rho_{ba}+V_{ac}\rho_{ca}-V_{bc}\rho_{ab}-V_{ca}\rho_{ac}\nonumber\\
i\dot{\rho}_{bb}&=&V_{ba}\rho_{ab}-V_{ab}\rho_{ba}\nonumber\\
i\dot{\rho}_{cc}&=&V_{ca}\rho_{ac}-V_{ac}\rho_{ca},\nonumber\\
\end{eqnarray}
which after rotating wave approximation RWA (i.e. after transforming-off the rapidly oscillating factors $\   $) with relaxation terms within the system, can be written as
\begin{eqnarray}\label{koncowe_0}
i\dot{\sigma}_{ab}&=&\Delta_1\sigma_{ab}-\Omega_1(\sigma_{bb}-\sigma_{aa})-\Omega_2\sigma_{cb}-i\gamma_{ab}\sigma_{ab}\nonumber\\
i\dot{\sigma}_{ac}&=&\Delta_2\sigma_{ac}-\Omega_1\sigma_{bc}-\Omega_2(\sigma_{cc}-\sigma_{aa})-i\gamma_{ac}\sigma_{ac}\nonumber\\
i\dot{\sigma}_{bc}&=&(\Delta_2-\Delta_1)\sigma_{bc}-\Omega_1^*\sigma_{ac}+\Omega_2\sigma_{ba}-i\gamma_{bc}\sigma_{bc}\nonumber\\
i\dot{\sigma}_{bb}&=&-\Omega_1^*\sigma_{ab}+\Omega_1\sigma_{ba}-i\Gamma_b\sigma_{bb}+i\Gamma_{cb}\sigma_{cc}\nonumber\\
i\dot{\sigma}_{cc}&=&-\Omega_2^*\sigma_{ac}+\Omega_2\sigma_{ca}-i\Gamma_c\sigma_{cc}-i\Gamma_{cb}\sigma_{cc},\nonumber\\
\end{eqnarray}
where following notation was introduced
\begin{eqnarray}
\Delta_1&=&E_a+\omega_1-E_b\nonumber\\
\Delta_2&=&E_a+\omega_2-E_c\nonumber\\
\Delta_2-\Delta_1&=&E_b-\omega_1-E_c+\omega_2+E_a-E_a\nonumber\\
E_a+\omega_1 &\approx& E_c\nonumber\\
E_a+\omega_2 &\approx& E_c
\end{eqnarray}
\begin{eqnarray}
V_{ab}&=&-\Omega_1e^{i\omega_1 t}\nonumber\\
V_{ac}&=&-\Omega_2e^{i\omega_2 t}\nonumber\\
\rho_{ab}&=&\sigma_{ab}e^{i\omega_1 t}\nonumber\\
\rho_{ac}&=&\sigma_{ac}e^{i\omega_2 t}\nonumber\\
\rho_{bc}&=&\sigma_{bc}e^{i(\omega_2-\omega_1) t}\nonumber\\
\gamma_{ab}&=&\frac{1}{2}\Gamma_b,\qquad \gamma_{ac}=\frac{1}{2}\Gamma_c\nonumber\\
\gamma_{bc}&=&\frac{1}{2}(\Gamma_b+\Gamma_c)\nonumber\\
\sigma_{aa}&=&1-\sigma_{bb}-\sigma_{cc}
\end{eqnarray}
Here, we assumed that there is no relaxation outside the system, so $\sum_{i} \sigma_{ii}=1$.
The propagation equation for signal fields 1 or 2 in slowly varying envelope approximation (SVEA) in the conditions of resonances reads
\begin{equation}\label{eq:propag1}
\frac{\partial \varepsilon_i}{\partial z}+\frac{\sqrt{\epsilon}}{c}\frac{\partial \varepsilon_i}{\partial t}=\frac{ik_i}{2\varepsilon_0\epsilon}P_i,
\end{equation}
where
 $P_i$ are the components of the polarization corresponding to the two excitonic resonances. Polarizations is determined by  the appropriate off-diagonal  elements of denstity matrix $\sigma_{ab}$ or $\sigma_{ac}$. For a weak signal pulse, the polarization in Fourier picture, for a given frequency, is proportional to the field and to an electric susceptibility $P_i(\omega)=\epsilon_0\chi_i(\omega)\varepsilon_i(\omega)$.

\begin{figure}[ht!]
\includegraphics[width=.7\linewidth]{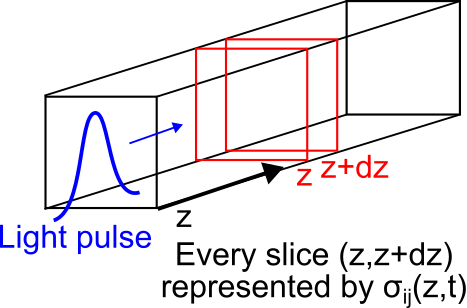}
\caption{Geometry of the considered system} \label{geometry1}
\end{figure}

Now, we will focus on a geometry depicted on Fig. \ref{geometry1}, where a crystal of a thickness   $L \sim 10 \mu$m  (length along $z$ axis) and macroscopic size along $x,y$ axes is considered. This makes the system quasi 1-dimensional, allowing one to use either the propagation equation (\ref{eq:propag1}) or a more general numerical approach  described in Appendix A.
According to Eq. (\ref{eq:propag1}) we consider a light pulse that  propagates through a Cu$_2$O crystal in the $z$ direction. The pulse is a linearly polarized plane wave, with a wave vector parallel to the $z$ axis. In order to simulate a propagation of such a pulse in the presence of excitons, two-steps calculations need to be performed:
\begin{itemize}
\item the set of equations (\ref{koncowe_0}) is solved for every position $z$. This means that individual values $\sigma_{ij}(z,t)$ represent exciton populations and coherences in a small volume (slice) of the crystal. By solving Eqs. (\ref{koncowe_0}), one obtains polarization $P(z,t)$. In case where two optical fields are considered, the total polarization is $P=P_1+P_2$,
\item Maxwell's equations are solved for a given incident field (laser pulse), using the calculated medium polarization. 
\end{itemize}
Both steps are done numerically, with finite time step $\Delta t$. First order Newton integration and second order FDTD method are used, correspondingly, see Appendix A for details. In this paper, we focus on the transmission coefficient of the crystal, which is calculated from the ratio of output to input electric field, e.g. $T=E_{max}(z=L)/E_{max}(z=0)$, where $E_{max}(z)$ is the maximum field value reached in the simulation time, at a given coordinate $z$, and $L$ is crystal length.

\subsection{Basic spectral and temporal properties of the pulse}
\begin{figure}[ht!]
a)\includegraphics[width=.95\linewidth]{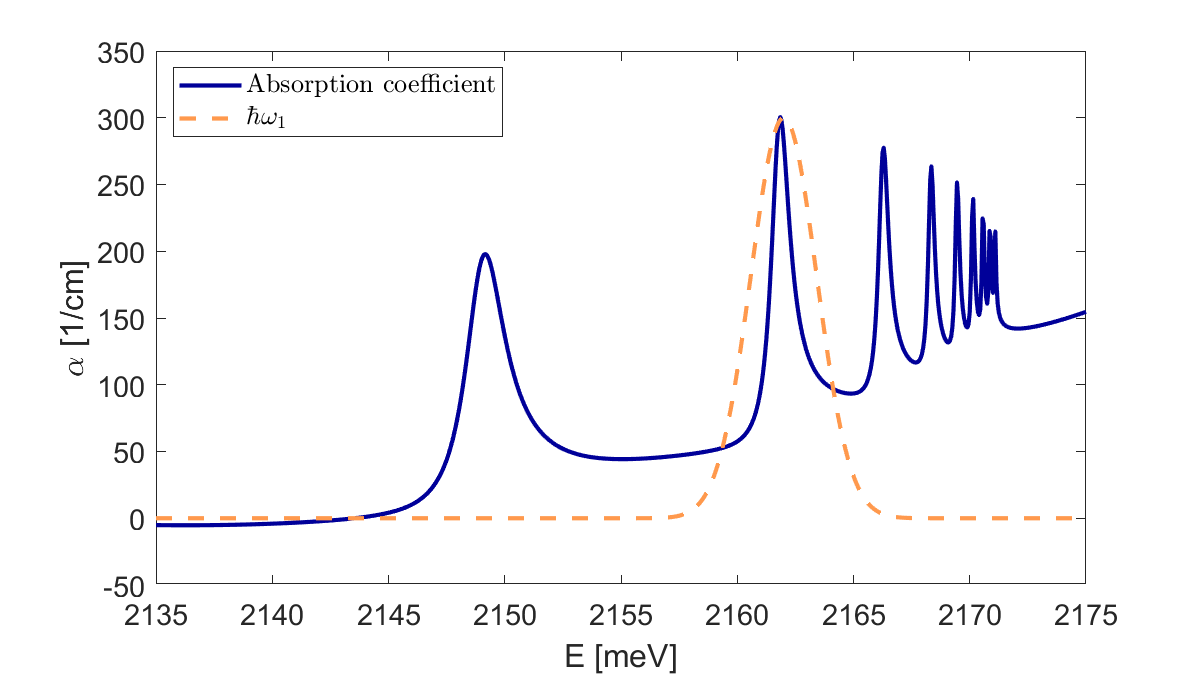}
b)\includegraphics[width=.95\linewidth]{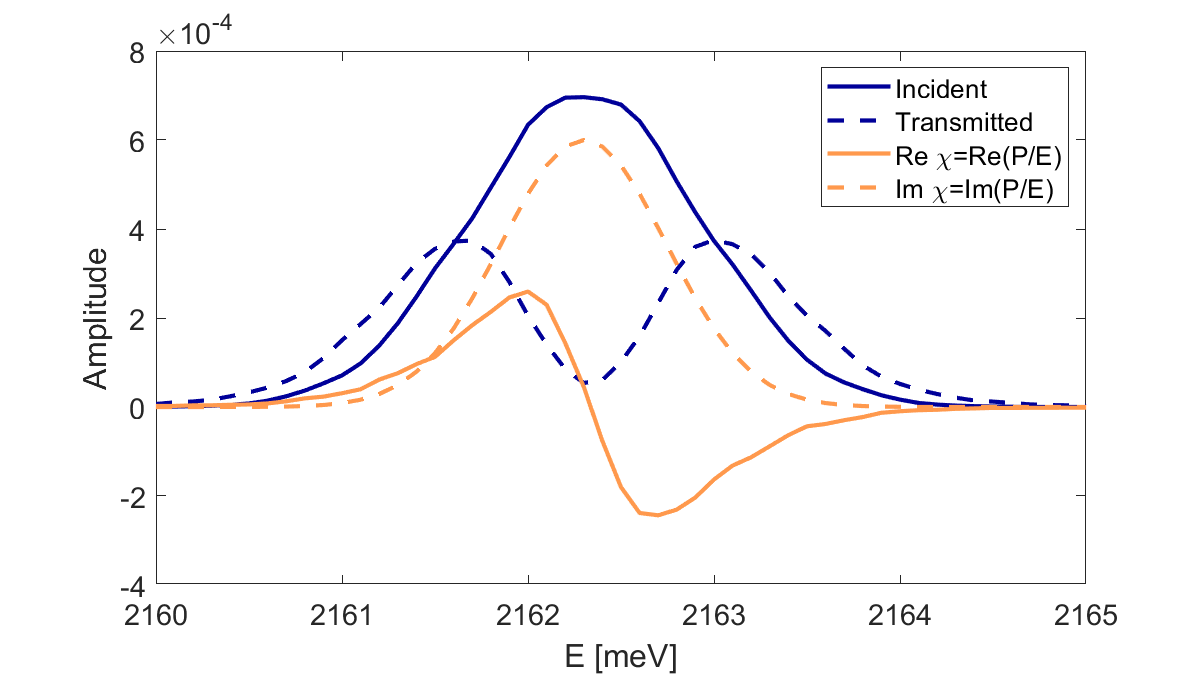}
\caption{a) absorption coefficient of Cu$_2$O crystal, exhibiting multiple excitonic peaks. Dashed line represents a Gaussian laser pulse. b) zoom on a single exciton resonance ($n$=3); numerically calculated frequency spectra of incident and transmitted field, as well as medium susceptibility are shown.}\label{fig:widmo}
\end{figure}

Following recent experiments \cite{Thomas}, we focus on the system dynamics when illuminated with short $\sim ps$ pulse. In the upcoming sections we will consider a single pulse with frequency $\hbar\omega_1$ and a linewidth that is comparable to the linewidth of the excitonic state, then in section III B propagation of two subsequent pulses will be analyzed. Two frequency pump-probe setup will be discussed later in section \ref{sek_pump}.\\ In particular here, we consider the case where the incident laser can reliably excite one lower excitonic state, but its linewidth is slightly wider than the excitonic resonance, as depicted on Fig. \ref{fig:widmo}.

One can see that in the presented example, the top panel of Fig. \ref{fig:widmo} depicts the absorption coefficient, given by
\begin{equation}
\alpha(\omega) = \frac{\omega}{c}Im~n(\omega),
\end{equation}
where $n(\omega)=\sqrt{\epsilon(\omega)}$ is the complex refraction index of the medium. We can separate the permittivity $\epsilon(\omega)$ into two parts $$\epsilon(\omega)=\epsilon_b+\chi(\omega),$$ where $\epsilon_b=7.5$ is the bulk permittivity and the susceptibility $\chi(\omega)$, which is in general a complex function, representing the excitonic contribution. The medium susceptibility  $\chi \sim \sigma_{aj}$ can be calculated solving stationary set of equations (\ref{koncowe_0});   its real part describes dispersion while an imaginary part absorption of the medium.

\begin{figure}[ht!]
\includegraphics[width=.95\linewidth]{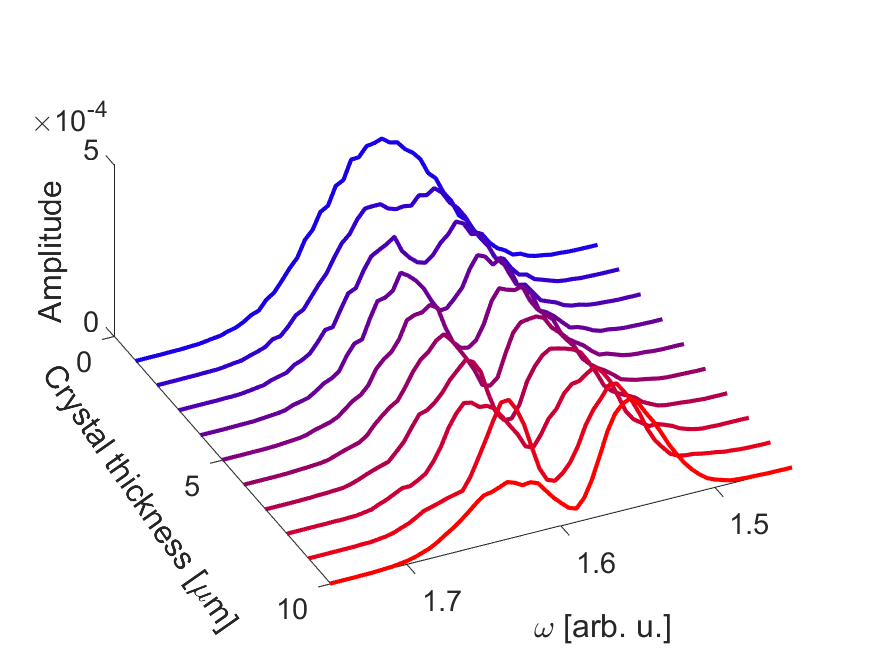}
\caption{The transmission spectrum of the crystal in a frequency range of a single excitonic resonance, as a function of crystal thickness.}\label{fig:widmo2}
\end{figure}

\begin{figure}[ht!]
\includegraphics[width=.95\linewidth]{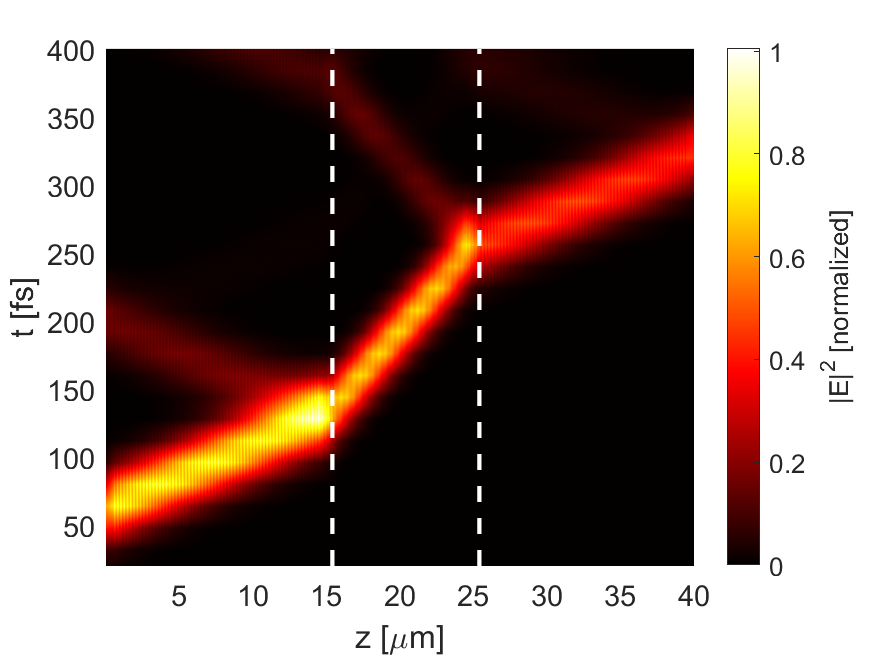}
\caption{Spatio-temporal distribution of the field illustrating the propagation of a single pulse through $L=10$ $\mu$m crystal (marked by white, dashed lines).}\label{fig:propag1}
\end{figure}

The laser field (dashed line on Fig. \ref{fig:widmo} a) is tuned to $n=3$ exciton resonance. The substantial linewidth of the laser pulse has a huge impact on propagation dynamics and transmission coefficient. Specifically, on the \ref{fig:widmo} b), example numerical results for a single field tuned to $n$=3 state are presented. The incident field is a Gaussian, but the transmitted field spectrum has a local minimum in the center. This is caused by the fact that the frequencies that are tuned perfectly to the exciton resonance are heavily attenuated due to the large value of the absorption coefficient $\sim Im~\chi$. On the other hand, the slightly detuned parts of the incident field propagate more freely. This phenomenon is further illustrated on Fig. \ref{fig:widmo2}, where a transmission spectrum is presented; it is calculated by applying a Fourier transform on the time-dependent transmission $T=E_{out}(t)/E_{in}(t)$, with input and output corresponding to electric field values immediately in front and behind the Cu$_2$O crystal, obtained from FDTD simulation as outlined in Appendix A. One can clearly see the formation of a dip in the center of the spectrum of the incident Gaussian pulse. Thus, for a sufficiently long crystal, the spectrum of the transmitted field will be characterized by two separate peaks corresponding to the wings of the original, incident pulse.
When discussing pulse propagation, it is important to mention the group velocity given by
\begin{equation}
V_g=\frac{c}{n+\omega\frac{\partial n}{\partial \omega}}.
\end{equation}
In the case in which the medium is characterized by normal dispersion $\frac{\partial n}{\partial \omega}>0$, the group velocity of the pulse will be slower than the phase velocity $c/n \sim 0.365$ c.  Fig. \ref{fig:propag1} illustrates such a case, with reduction of group velocity as the pulse enters the crystal.
Outside the crystal, the pulse velocity (slope of the curve) is $V_g \approx c \approx 0.3$ $\mu$m/fs. Inside, it slows down to $V_g \sim 0.1$ $\mu$m/fs. Furthermore, one can see the weak reflection streaks on the outer and inner surface of the crystal occurring at $t \sim 120$ fs and $t \sim 250$ fs.

Now, let's consider propagation of a spectrally wide pulse.
In the spectral area of the excitonic resonance, one has so-called anomalous dispersion $\partial \chi/\partial \omega < 0$, clearly visible on Fig. \ref{fig:widmo}, bottom panel. This means that the group velocity of a pulse tuned to the excitonic resonance is slightly faster than the phase velocity $V_p \approx c/\sqrt{7.5} = 0.365$ c. Naturally, such a pulse is also heavily damped, particularly close to the resonance. For a spectrally wide pulse considered here, its central part corresponds to this anomalous dispersion region, while left and right wings will propagate under the conditions of normal dispersion $\partial \chi/\partial \omega > 0$ and smaller absorption. Thus, at the given sufficient propagation distance, one can observe a splitting of the pulse into "slower" and weaker "faster" part, which comes from  wings of both part of the susceptibility. This is further discussed in Appendix A.

\subsection{The Rydberg blockade}\label{sekcja_rydb}
A characteristic feature of  systems with Rydberg excitons stems from their strong mutual interactions, which due to their dimension $\sim n^2$, become important on length scales  much larger then the distance between excitons. This leads to so-called exciton Rydberg  blockade  that prevents the optical excitation of two excitons within typical distances of several $\mu$m.  Under such conditions, the relevant interactions are modified by direct Coulomb interaction between excitons. It should be stressed that particularly for excitonic states with a high principal quantum number $n$ the influence of the blockade is considerable since the coefficient $C_6 \sim n^{11}$, so that the van der Waals interaction Eq.(5) leads to a significant shift of exciton energy levels.

Rydberg blockade  complicates the system dynamics which has to be treated by extending above single-body approach and  to consider the ensemble composed of N interacting three-level sub-systems at position $r_i$. However, the number of created excitons depends on the laser intensity, therefore for lower exciton states created in laser fields of  low  intensities the blockade effect can be negligible, while for high excitonic states Rydberg blockade may be significant even for a very low laser power.

The presence of Rydberg-Rydberg interactions results in a strongly correlated many-body dynamics.
 
Discussing approximate description of the excitation dynamics at moderate excitons densities we consider
the time evolution of the reduced density matrix elements, incorporating projection operators of the type $\sigma^{(i)}_{pq}= |p\rangle \langle q|$ ($p,q=a,b,c$) \cite{Sevincli}.
\\\\
Including the influence of Rydberg blockade, the set of Eqs. (\ref{koncowe_0}) is modified as follows 
\begin{eqnarray}\label{sigma_koncowe}
i\dot{\sigma}_{ab}^{(i)}&=&\Delta_1\sigma_{ab}^{(i)}-\Omega_1(\sigma_{bb}^{(i)}-\sigma_{aa}^{(i)})-\Omega_2\sigma_{cb}^{(i)}-i\gamma_{ab}\sigma_{ab}^{(i)}\nonumber\\&&-\sum_{i\neq j} V_{ij}\sigma_{bb}^{(j)}\sigma_{ba}^{(i)}\nonumber\\
i\dot{\sigma}_{ac}^{(i)}&=&\Delta_2\sigma_{ac}^{(i)}-\Omega_1\sigma_{bc}^{(i)}-\Omega_2(\sigma_{cc}^{(i)}-\sigma_{aa}^{(i)})-i\gamma_{ac}\sigma_{ac}^{(i)}\nonumber\\&&-\sum_{i\neq j} V_{ij}\sigma_{cc}^{(j)}\sigma_{ca}^{(i)}\nonumber\\
i\dot{\sigma}_{bc}^{(i)}&=&(\Delta_2-\Delta_1)\sigma_{bc}^{(i)}-\Omega_1^*\sigma_{ac}^{(i)}+\Omega_2\sigma_{ba}^{(i)}-i\gamma_{bc}\sigma_{bc}^{(i)}\nonumber\\&&-\sum_{i\neq j} V^{*}_{ij}\sigma_{cc}^{(j)}\sigma_{bc}^{(i)}\nonumber\\
i\dot{\sigma}_{bb}^{(i)}&=&-\Omega_1^*\sigma_{ab}^{(i)}+\Omega_1\sigma_{ba}^{(i)}-i\Gamma_b\sigma_{bb}^{(i)}+i\Gamma_{cb}\sigma_{cc}^{(i)}\nonumber\\
i\dot{\sigma}_{cc}^{(i)}&=&-\Omega_2^*\sigma_{ac}^{(i)}+\Omega_2\sigma_{ca}^{(i)}-i\Gamma_c\sigma_{cc}^{(i)}-i\Gamma_{cb}\sigma_{cc}^{(i)},\nonumber\\
\end{eqnarray}
where Rydberg excitons mutually interact via isotropic  potential $V_{ij}=\frac{C_6}{r^6_{ij}}$.

The resulting dynamical equations for a single-body elements involve two-body elements, whose dynamics requires knowledge of three-body elements etc. The chain can be truncated at an appropriate step depending on the number of interacting excitons. However, this general approach is tremendously demanding numerically.

An alternative method suitable to describe the case with arbitrary interaction strengths  is so-called Monte Carlo approach \cite{Morawetz2023}, where we calculate the average Rydberg blockade energy shift and its standard deviation as a function of exciton density, under the assumption that excitons are distributed randomly. 

For the interstate van der Waals potential  $V^{}_{ij}(n,n')$, we adapt the model of the van der Waals interaction used for Rydberg excitons in \cite{Heckotter21}, where the $C_6$ constant for two excitons characterized with principal quantum numbers $n_1$, $n_2$ is given by
\begin{equation}\label{eq:C6}
C_6 \sim \frac{n_1^4n_2^4}{n_1^{-3}+n_2^{-3}}.
\end{equation}
This results in $n^{11}$ scaling \cite{Kazimierczuk} when a single excitonic state $n_1=n_2=n$ is present. For the exact value of $C_6$, we use the estimation value  $C_6=0.2$ meV$\mu m^6$ for n=16 calculated by Morawetz \textit{et al} \cite{Morawetz2023}.

The Monte Carlo procedure is performed as follows:
\begin{itemize} 
\item a test volume of 1000 $\mu$m$^3$ is defined. Within this volume, individual excitons are placed at random, uniformly distributed positions. For each added exciton, its interaction with all previous excitons is calculated as
\begin{equation}
\Delta E_{VdW} = \sum \frac{C_6}{r^6},
\end{equation}
where summation is performed over all previously put excitons. The resulting energy and corresponding exciton density are recorded, \\
\item  repeating the procedure multiple times, sufficient number of energy values can be obtained to reliably estimate the mean energy and its standard deviation.\end{itemize}

\begin{figure}[ht!]
a)\includegraphics[width=.9\linewidth]{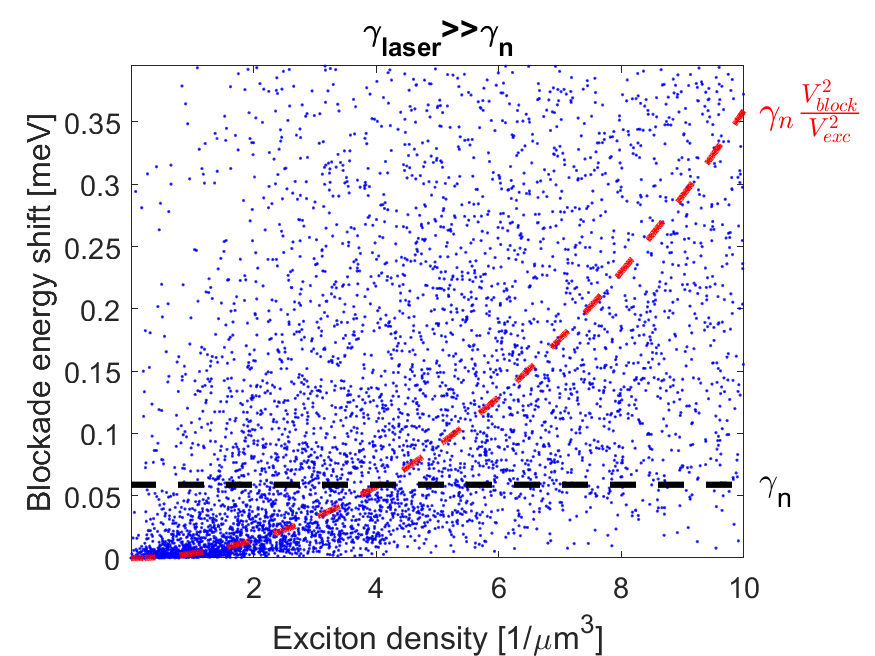}
b)\includegraphics[width=.9\linewidth]{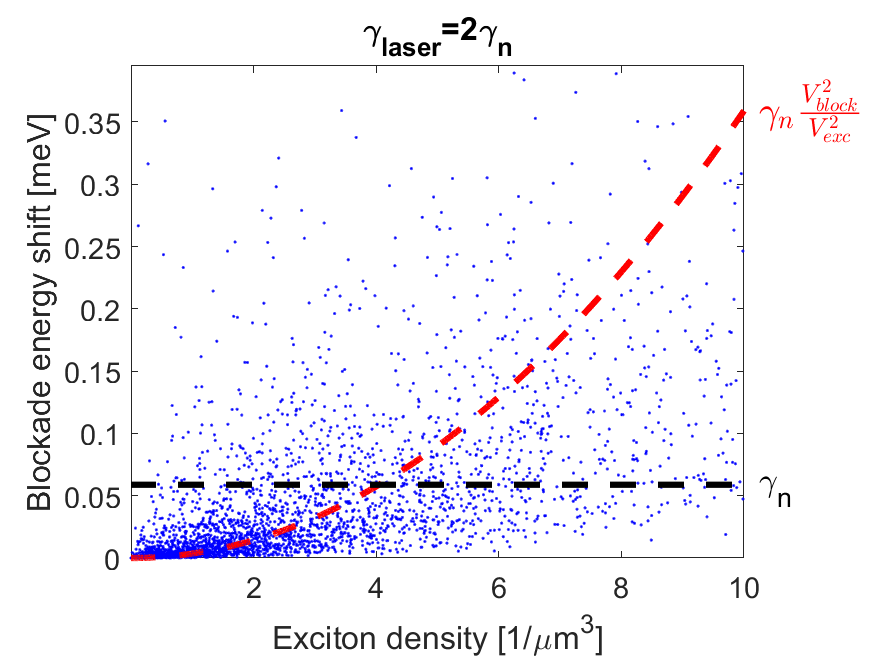}
\caption{Results of the Monte Carlo simulation for n=7 excitons created by a) spectrally wide and b) spectrally narrow light source.} \label{fig:monte}
\end{figure}

An important consideration is the probability of the exciton creation depending on the energy shift. Fig. \ref{fig:monte} a) depicts the case where the illuminating field is spectrally very wide. This means that the exciton can be created at any position, regardless of the energy shift caused by van der Waals interaction.

As expected, the energy shift has a considerable random spread and a general tendency to increase with increasing exciton density. Specifically, the $n=7$ exciton considered here is characterized by a blockade volume of about $V_{block}\approx0.25$ $\mu$m$^3$ and corresponding density of $4/\mu$m$^3$. The $r^{-6}$ scaling of interaction energy implies that one can approximate the results with a relation $\sim 1/V_{exc}^2$, where $V_{exc}$ is the volume per exciton (inverse of the density). This relation is shown by a red, dashed line on Fig. \ref{fig:monte}. Crucially, when the volume per exciton matches the blockade volume, the energy shift becomes equal to exciton linewidth $\gamma_n$ (black, dashed line).

 Fig. \ref{fig:monte}b) depicts calculation results for a light source with limited spectral width. Specifically, we assume that the shape of a laser pulse is a Gaussian with linewidth $\gamma_{laser}=2\gamma_{n}$. In this case, after random position is chosen in the test volume and the corresponding energy $\Delta E_{VdW}$ is calculated, the exciton is placed in that position with a finite probability $p \sim \exp[-\Delta E_{VdW}/(\gamma_{laser}+\gamma_n)]$. Thus, very high energies become less probable. This is reflected on Fig. \ref{fig:monte} b), where considerably smaller number of points is visible near the top of the plot. The dependence of the energy on the density becomes more linear and the standard deviation of energy  decreases. The exclusion of high blockade interaction energy excitons corresponds to a more spatially uniform distribution, with no cases of excitons very close to each other. A detailed analysis of the spatial considerations is presented in \cite{Morawetz2023}, where the reduction of the probability of exciton  creation is calculated directly and the corresponding reduction of absorption is compared to recent experimental study \cite{nature2021}. In contrast to \cite{Morawetz2023}, here we focus on the energy $\Delta E_{VdW}$ that results in a detuning from the laser field which, in turn, indirectly affects the absorption.

As mentioned above, we consider isotropic van der Waals potential for mutual exciton interaction. However, in the quasi one-dimensional, mesoscopic model shown on Fig. \ref{geometry1}, individual excitons are not tracked explicitly. Rather, one slice corresponding to the position $(z,z+dz)$ is characterized by a set of values of $\sigma_{ij}$. In a numerical representation, with a finite value of $dz$, such a slice represents some small volume with a certain number of excitons inside it. Based on the number of excitons, the slice is characterized with some average value of van der Waals interaction $\Delta E_{VdW}(n_1,n_2,\sigma_{bb},\sigma_{cc})$ and standard deviation $\sigma_{VdW}(n_1,n_2,\sigma_{bb},\sigma_{cc})$. After estimating the above parameters for all exciton densities obtained within simulation, one can use a heuristic approach, where at each time step, for each slice $(z,z+dz)$, a random value of blockade energy is attributed corresponding to $\Delta E_{VdW}(z,t)$ and $\sigma_{VdW}(z,t)$.

\section{Simulation results}

\subsection{Optical bleaching}
\begin{figure}[ht!]
a)\includegraphics[width=.8\linewidth]{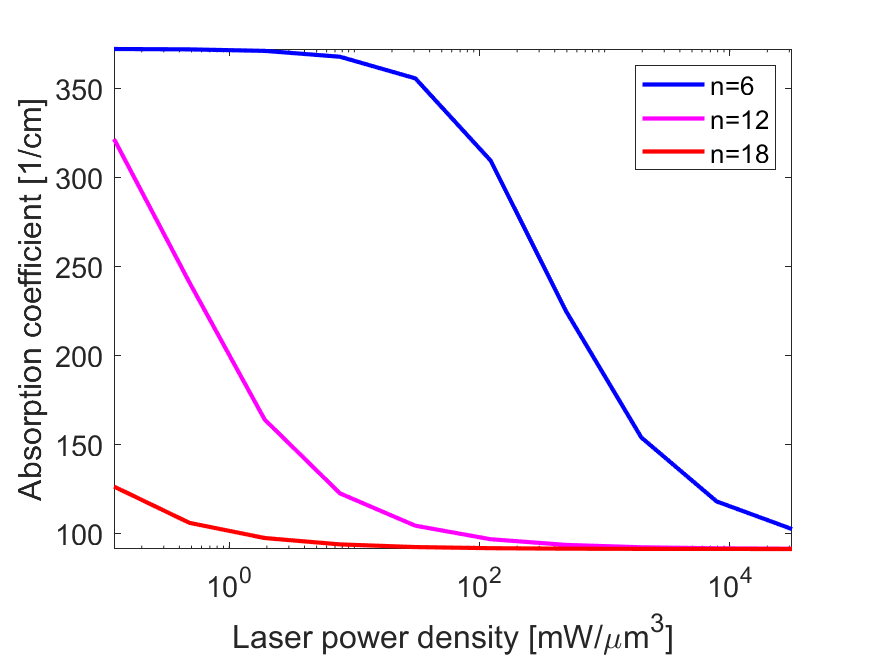}
b)\includegraphics[width=.8\linewidth]{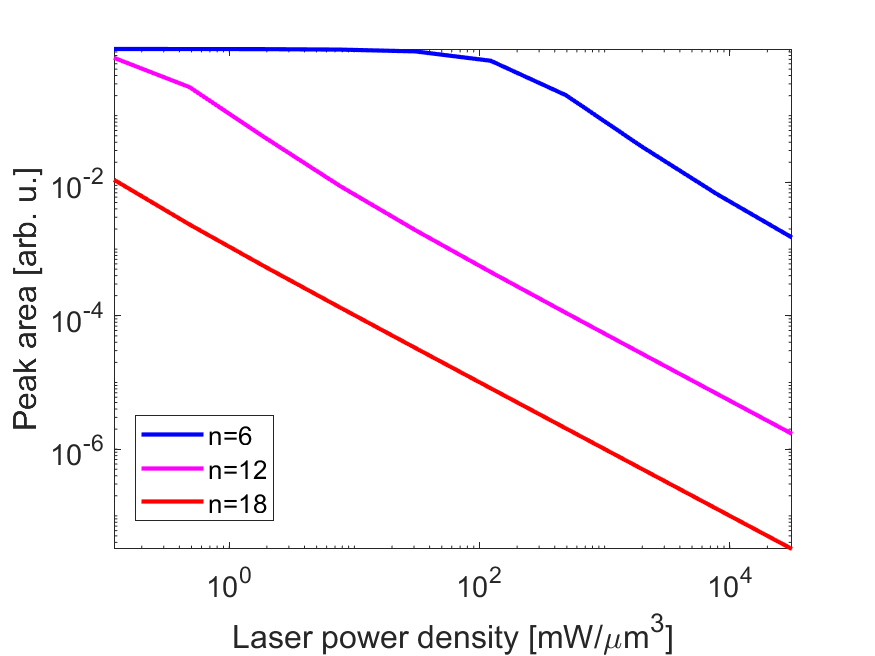}
\caption{a) Absorption coefficient as a function of laser power, calculated for three excitonic states. b) Oscillator strength as a function of laser power.}\label{fig:verif1}
\end{figure}

For the purpose of analyzing the transmission of laser pulses through the system, the most important effect of the blockade is its power-dependent impact on the absorption coefficient (so-called bleaching). In order to validate our approach, we refer to experimentally measured absorption coefficients \cite{Malerba} and recent Monte Carlo many-body simulations \cite{Morawetz2023}. First, we will calculate the transmission and then the absorption coefficient of the system by performing a simulation with continuous, monochromatic incident field tuned to a specified excitonic resonance. Results are shown on Fig. \ref{fig:verif1} a). For $n$=6, the absorption coefficient keeps a fairly constant value $\sim 300$/cm until some threshold power is reached and the system saturates. In other words, the exciton density reaches a value where the Rydberg blockade induced energy shift exceeds the excitonic linewidth, preventing creation of further excitons and thus suppressing absorption. For $n$=12 and $n$=18, the smallest considered power density is already within saturation range. It should be pointed out that this bleaching mechanism affects excitonic part of the absorption, but not absorption background not related to the excitons. In our model we include background absorption of the order of $\sim 80$/cm, which is independent of excitons, to more closely represent typical experimental conditions \cite{Malerba}. Analogously to the measurements done  by Kazimierczuk \emph{et al} \cite{Kazimierczuk}, we can calculate the area of excitonic absorption peak, which is proportional to effective oscillator strength of the resonance. The results, shown on Fig. \ref{fig:verif1} b), are consistent with experimental data presented in  \cite{Kazimierczuk}.

\begin{figure}[ht!]
\includegraphics[width=.8\linewidth]{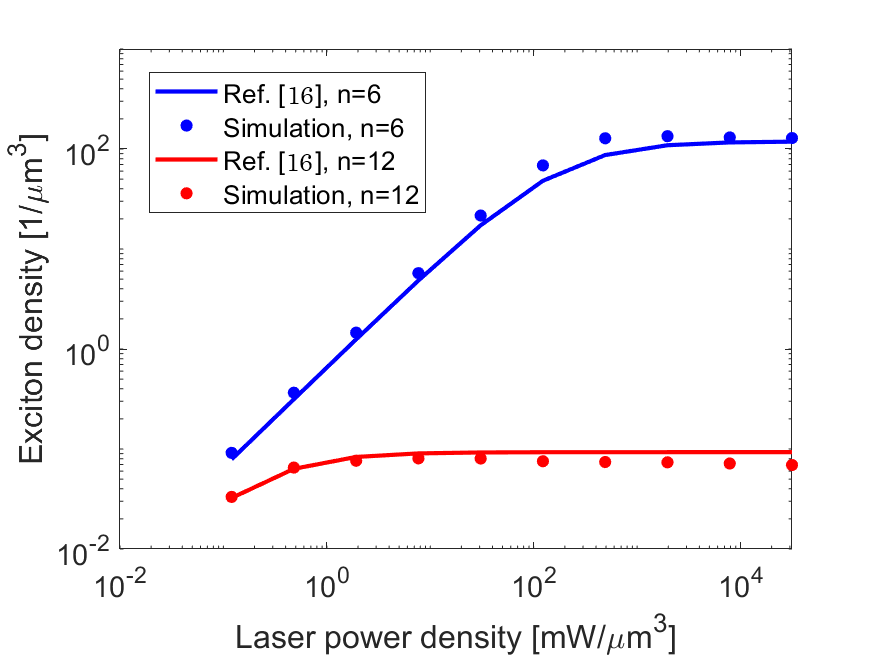}
\caption{Exciton density as a function of power, compared with results from \cite{Morawetz2023}.}\label{fig:verif2}
\end{figure}

Another useful quantity that can be extracted directly from the simulation results is the exciton density, defined as $\rho_{exc}=\sigma_{bb}+\sigma_{cc}$. Again, we use continuous, monochromatic field and record the density as a function of power. Results for $n$=6 and $n$=12 are shown on Fig. \ref{fig:verif2}. As a verification, we compare our results with the formula proposed in \cite{Morawetz2023}, obtaining an excellent agreement. Like in the absorption results, the saturation of the system at high power is evident. Due to the rapid Rydberg blockade scaling $\sim n^{11}$, system with  higher excitons system saturates  earlier. It should be stressed that the considered power densities are excessive in a continuous illumination scenario, but easily achieved for sub-ps pulses. Thus, even for relatively low n states, Rydberg blockade considerations are highly relevant.

\begin{figure}[ht!]
\includegraphics[width=.8\linewidth]{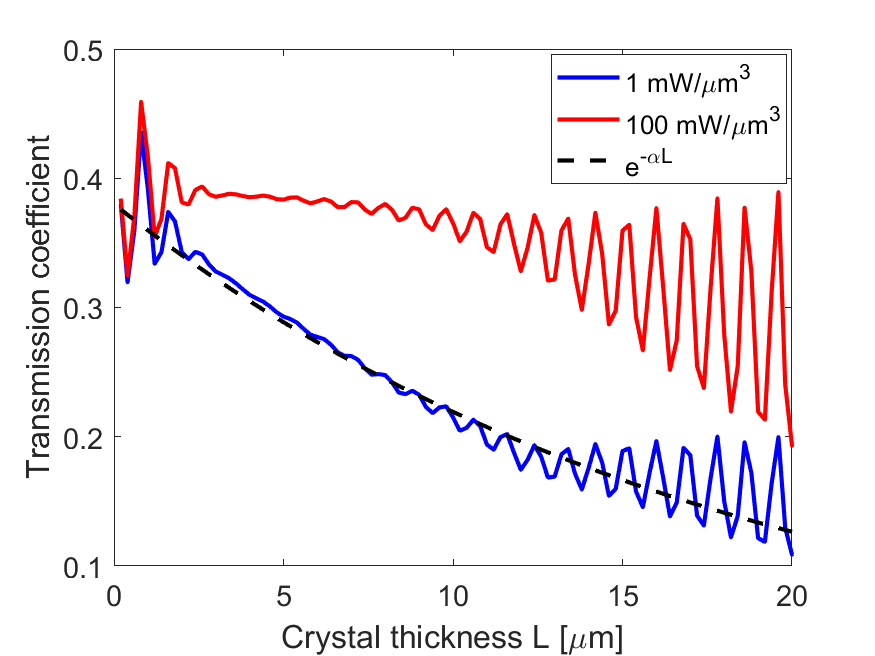}
\caption{Transmission coefficient as a function of crystal length, for two laser powers. Calculated for a 100 fs pulse tuned to $n=6$ exciton resonance.}\label{fig:trans_pow}
\end{figure}

A further insight into the bleaching process can be gained by considering the spatial distribution of the field. As the pulse propagates into the medium, its amplitude is reduced due to the absorption. Thus, the power absorbed by the crystal is reducing exponentially away from the surface. As a result, the exciton density, which is proportional to the absorbed energy density, will also be nonuniform. This leads to position-dependent blockade effects such as bleaching. The effects of this uneven distribution are visible on Fig. \ref{fig:trans_pow}.
In the low power regime, the transmission follows the standard exponential decay. However, one can see significant oscillations of the transmission coefficient, that can be attributed to Fabry-Perot type interference. As mentioned in the previous section, a spectrally wide pulse has multiple frequency components characterized by slightly varying wavelengths. This results in a more complex interference pattern than in the case of a monochromatic field. For the high power pulse, the transmission stays at a roughly constant value up to $L \sim 10$ $\mu$m, for which the system saturates, so that $\alpha \rightarrow 0$. However, for a thicker crystal, the energy contained in the pulse is insufficient to fully saturate the system. As a result, transmission coefficient starts to decrease for $L>10$ $\mu$m.

\subsection{Two pulses}
\begin{figure}[ht!]
\includegraphics[width=.9\linewidth]{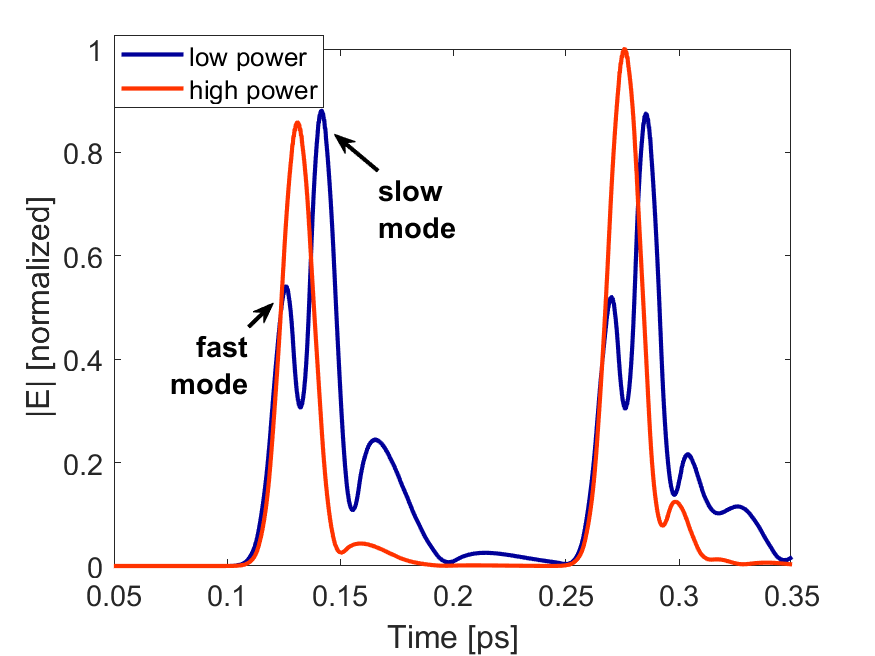}
\caption{The time evolution of the field exiting the crystal in the case of two consecutive input pulses.}\label{fig:two1}
\end{figure}

The first step in the study of the dynamic properties of the Rydberg blockade is to consider two consecutive, identical pulses propagating through the crystal. The first pulse generates some initial exciton population, which in turn affects the propagation conditions for the second pulse. The results of a simulation of such system are shown on Fig. \ref{fig:two1}, where the normalized electric field of the pulse exiting the crystal is shown.
Both pulses following each other were tuned to $n$=10 exciton resonance. Two distinct cases for low and high power pulses have been tested. 

The low power pulse was weak enough for the blockade effects to be negligible. In such a case, one can expect strong absorption and dispersion. This is indeed the case here. On Fig. \ref{fig:two1}, the Gaussian pulse has clearly split into two distinct parts; the first peak at $t \sim 0.12$ ps is the faster mode corresponding to the part of the spectrum characterized by strong absorption and anomalous dispersion. Slightly later, the second peak of the slower mode is present. This peak is higher due to the lower absorption on the wings of the exciton absorption line. The second pulse, starting at $t \sim 0.25$ ps, is identical to the first one. The low exciton density created by the first pulse was insufficient to affect the second pulse in meaningful way. The only noticeable difference between two low power pulses is their "tail" which represents the field exiting the crystal after multiple reflections inside. This part of the first pulse is still present in the crystal when the second pulse propagates through it.

The system dynamics is different for a high power pulse, which was sufficiently strong to saturate the medium. First, one can see that the pulse is no longer split. When the system is saturated with excitons, the effective oscillator strength of the excitonic resonance is reduced, as indicated on Fig. \ref{fig:verif1} a) and b). This means that both imaginary and real parts of susceptibility are greatly reduced. Thus, not only the absorption is suppressed, but also there is no noticeable dispersion. As a result, the whole pulse propagates at a group velocity approximately equal to the phase velocity. One can see that the high power peak is located between the fast (anomalous dispersion) and slow (normal dispersion) ones of the weak pulse. Another visible feature is the fact that the amplitude (height of the peak on Fig. \ref{fig:two1}) of the second high power pulse is higher than the first one. This is a consequence of the fact that the first pulse left a significant exciton population behind, which made the medium more transparent to the second pulse.  This phenomenon opens up a potential avenue of a study of the dynamic properties of excitons; the precise change of transmission coefficient between first and second pulse depends not only on pulse power, but also there is  an interplay between time period among pulses and lifetime of the excited state.

\subsection{Pump-probe setup}\label{sek_pump}

In this section, we consider a pump-probe setup with two distinct pulses characterized by frequencies $\omega_1$ and $\omega_2$. The first pulse $\hbar\omega_2$ is a pump field that excites a higher excitonic state $n=10$. The second probe pulse $\hbar\omega_1$ is tuned to a low exciton state $n=5$. The energy level schematic is depicted on Fig. \ref{fig:drabinka}. The weak probe pulse follows immediately after the pump pulse. Due to this technique, the pulsed excitation drives the system into a high-density regime beyond that is typically achieved in a CW experiment. Thus, it is expected that the pump and probe excitons will interact strongly through Rydberg blockade, according to Eq. (\ref{eq:C6}).

\begin{figure}[ht!]
\includegraphics[width=.9\linewidth]{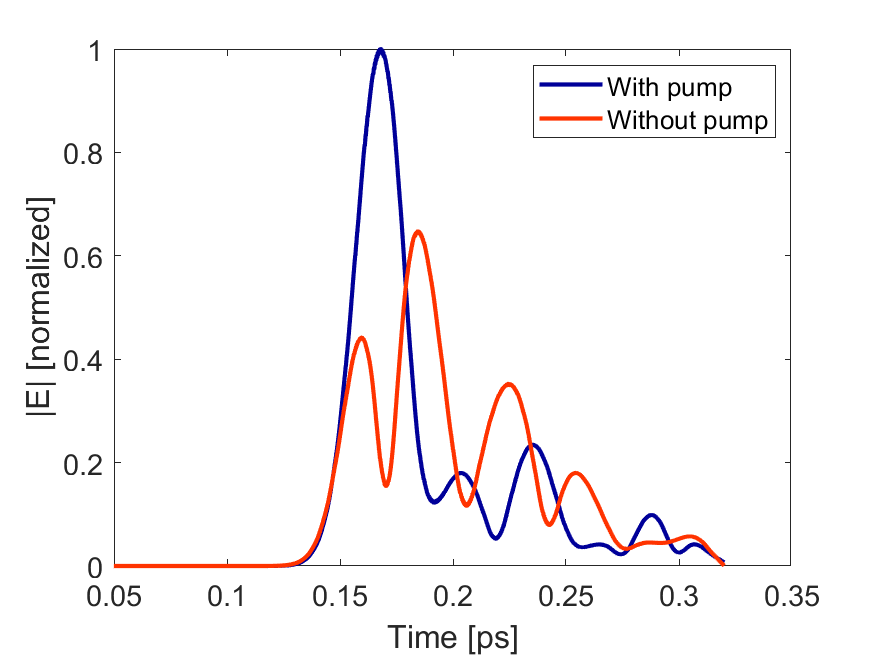}
\caption{The time evolution of the field exiting the crystal in the case of two simultaneous input pulses with different frequencies.}\label{fig:two2}
\end{figure}

The output probe field of such a setup is shown on Fig. \ref{fig:two2}. 
With no pump present, the output resembles the case shown on Fig. \ref{fig:two1}; the pulse  splits into two modes, with following later some additional peaks corresponding to the reflections in the crystal. In the case when pump is present, the splitting is suppressed and the transmission coefficient of the probe field is significantly higher. As mentioned above, we assume that the pump and probe pulses are nearly simultaneous; however, by introducing a delay of the probe pulse, one could gain some additional insight into the dynamics of the system. Such an approach would be a odd parity ($P,F$) exciton counterpart to the recent difference frequency generation (DFG) study of even parity ($S,D$) excitons \cite{Farenbruch2025}.

\begin{figure}[ht!]
\includegraphics[width=.9\linewidth]{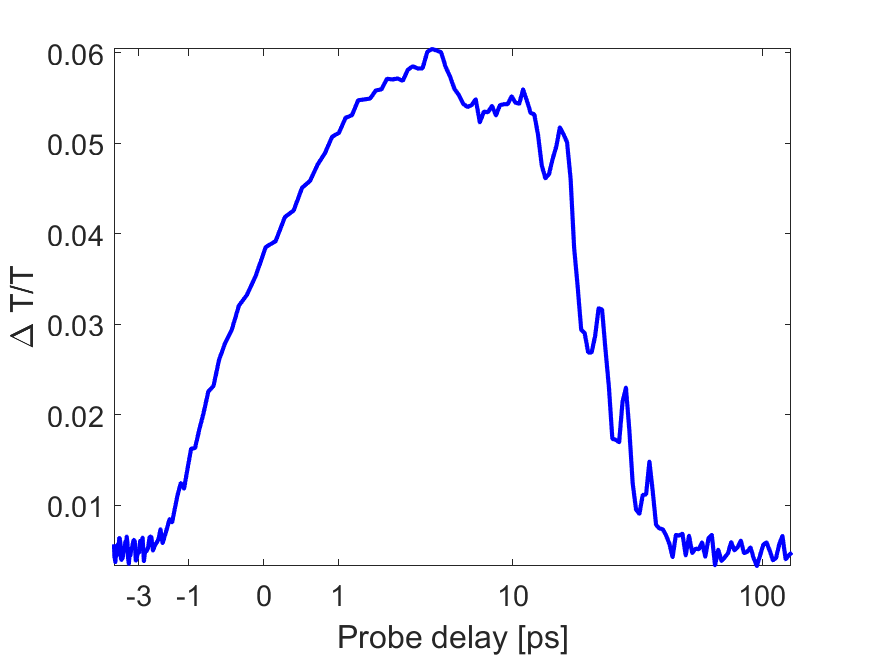}
\caption{Relative change of transmission of the probe pulse, as a function of the probe pulse delay. }\label{fig:two3}
\end{figure}

The Fig. \ref{fig:two3} depicts a relative change of probe transmission as function of probe pulse delay. The pump is tuned to $n$=7 exciton and the probe is tuned to $n$=6 state to compare our theoretical results with experimental data \cite{Minarik}. One can observe that the maximum change of transmission occurs for a small, positive delay $\sim 3$ ps. This corresponds to the situation where the probe pulse enters the medium immediately after the pump has been fully absorbed, so that the exciton population has reached its maximum. For larger delays, the population decays at a rate corresponding to the lifetime of the excited state $\sim 20$ ps. The results exhibit a good agreement with recent measurements reported in \cite{Minarik}. 
A characteristic feature of transmission on Fig. \ref{fig:two3} is the presence of two types of oscillations. On the left, rising part of the curve one can notice small oscillations with a period of $\sim 100$ fs. This corresponds to the propagation time through the crystal. Thus, the cause of these oscillations is most likely an interference between incident and reflected parts of the pulse. On the right side of the plot, stronger oscillations with period $\tau \sim 4$ ps are present. This period matches the energy spacing between $7P$ and $6P$ states i.e., $\Delta E=\hbar\frac{2\pi}{\tau} \sim 1$ meV. Thus, these oscillations can be attributed to coherent oscillations of exciton populations, similarly to the quantum beats  in emission spectra, which were recently observed and discussed theoretically  \cite{Thomas,myQB}.

The maximum relative change of transmission coefficient on Fig. \ref{fig:two3} is of the order of 6$\%$, which corresponds to the change of absorption coefficient of approximately $\sim 2/cm$. This matches the recent time-resolved pump-probe measurements reported in \cite{Panda}. Crucially, here we focus on short, high power pulses that are not subject to so-called purification process which happens in low power regime \cite{Panda} and exhibits relaxation at a microsecond timescale.

\section{Conclusions}
We have investigated the propagation  of short laser pulses in Rydberg exciton medium, which became a topical problem in a view of recent experiments in developing solid-state Rydberg optics.
By using a combination of density matrix approach, Monte Carlo simulations of the Rydberg blockade and FDTD simulations, we have numerically studied the propagation of short pulses through Cu$_2$O crystal, in the regime of strong Rydberg blockade. First, we have analyzed a pulse propagation through a  two-level system with ground state and a single excitonic state; in addition to the demonstration of the so-called optical bleaching, which depends on saturation effects and is a consequence of Rydberg blockade. We also have pointed out some unique potential effects such as splitting of the pulse that may occur for a sufficiently spectrally wide incident field. The interplay between the exciton lifetime and transmission of consecutive pulses is discussed. Finally, we have analyzed a three-level pump-probe setup and compared our results to recent experiments \cite{Panda,Minarik}, underscoring a good agreement between measured and calculated spectra.

\section{Acknowledgment}
Support from National Science Centre, Poland (Project No. OPUS 2025/57/B/ST3/00334), is greatly acknowledged.

\section{Appendix A: numerical calculations}
In order to study the propagation of the pulse inside the Cu$_2$O crystal, the time evolution of the electromagnetic field is calculated in a set of discrete positions $N \Delta z$, $N \in \mathbb{N}$. In this paper, we assume $\Delta z=1$ nm, which is a sufficiently small step to accurately represent the spatial distribution of the field considering that the typical wavelength inside Cu$_2$O is $\lambda \sim 200$ nm. The evolution is calculated in discrete time steps as well, with $\Delta t=0.1$ fs, which fulfills the Courant stability criterion $c\Delta t/\Delta z < 1$ \cite{Yee}.

The starting point of the calculations is the discretized version of Eqs. (\ref{sigma_koncowe}) in the form
\begin{eqnarray}\label{sigma_num}
{\sigma}_{ab}^{(t+\Delta t)}&=&{\sigma}_{ab}^{(t)}+i\Delta t \left[-\Delta_1\sigma_{ab}^{(t)}+\Omega_1(\sigma_{bb}^{(t)}-i\sigma_{aa}^{(t)})+\Omega_2\sigma_{cb}^{(t)}\right.\nonumber\\&&-\left. i\gamma_{ab}\sigma_{ab}^{(t)}+\Delta E_{VdW,ab}^{(t)}\right],\nonumber\\
{\sigma}_{ac}^{(t+\Delta t)}&=&{\sigma}_{ac}^{(t)}+i\Delta t\left[-\Delta_2\sigma_{ac}^{(t)}+\Omega_1\sigma_{bc}^{(t)}+\Omega_2(\sigma_{cc}^{(t)}-i\sigma_{aa}^{(t)})\right.\nonumber\\&&-\left. i\gamma_{ac}\sigma_{ac}^{(t)}+\Delta E_{VdW,ac}^{(t)}\right],\nonumber\\
{\sigma}_{bc}^{(t+\Delta t)}&=&{\sigma}_{bc}^{(t)}+i\Delta t\left[-(\Delta_2-\Delta_1)\sigma_{bc}^{(t)}+\Omega_1^*\sigma_{ac}^{(t)}-\Omega_2\sigma_{ba}^{(t)}\right.\nonumber\\  &&- \left. i\gamma_{bc}\sigma_{bc}^{(t)}+\Delta E_{VdW,bc}^{(t)}\right],\nonumber\\
{\sigma}_{bb}^{(t+\Delta t)}&=&{\sigma}_{bb}^{(t)}+i\Delta t\left[\Omega_1^*\sigma_{ab}^{(t)}-\Omega_1\sigma_{ba}^{(t)}-i\Gamma_b\sigma_{bb}^{(t)}+i\Gamma_{cb}\sigma_{cc}^{(t)}\right],\nonumber\\
{\sigma}_{cc}^{(t+\Delta t)}&=&{\sigma}_{cc}^{(t)}+i\Delta t\left[\Omega_2^*\sigma_{ac}^{(t)}-\Omega_2\sigma_{ca}^{(t)}-i\Gamma_c\sigma_{cc}^{(t)}-i\Gamma_{cb}\sigma_{cc}^{(t)}\right],\nonumber\\
\end{eqnarray}
where $\Delta E_{VdW,ij}^{(t)}$ is the corresponding Van der Waals interaction energy obtained from Monte Carlo simulation, dependent on the current populations of excitons $\sigma_{bb}^{(t)}$, $\sigma_{cc}^{(t)}$. The above system of equations is a first order Newton iteration with a finite step $\Delta t$; the choice of the first order method is dictated by the fact that the population dynamics (changes of $\sigma_{ij}$) are slow compared to the changes of electric field at optical frequencies.

After the calculation of the new populations $\sigma_{ij}^{(t+\Delta t)}$ from current ones $\sigma_{ij}^{(t)}$, new values of the Van der Waals interaction energy $\Delta E_{VdW,ij}^{(t+\Delta t)}$ are estimated as described in section \ref{sekcja_rydb}. Note that the above calculation requires also the current state of the electric field through the terms $\Omega_1$, $\Omega_2$, according to Eq. (\ref{eq:omegaj}). The dipole moments present in that equation are calculated from first principles as described in \cite{myQB}.

The next step of the calculation is based on the Finite Difference Time Domain (FDTD) method. This second-order numerical approach is suitable for the calculation of the rapidly changing optical field and allows one to simulate the propagation of arbitrary fields. In the 1-dimensional case considered here, we define a grid of the electric field values $E_x(N \Delta z)$ and magnetic field values $H_y(N \Delta z)$ where without a loss of generality, for a plane wave propagating in the z direction, it is assumed that $\vec E=[E_x,0,0]$, $\vec H=[0,H_y,0]$. From the Maxwell's equations, one can derive a set of discretized evolution equations
\begin{eqnarray}\label{FDTD}
H_{z+\frac{1}{2}}^{t+\frac{1}{2}} &=& H_{z+\frac{1}{2}}^{t-\frac{1}{2}} - \frac{\Delta t}{\mu_0} \frac{E_{z+1}^{t} - E_{z}^{t}}{\Delta z},\nonumber\\
E_z^{t+1} &=& E_{z}^{t} -\frac{\Delta t}{\epsilon_0}\frac{H_{z+\frac{1}{2}}^{t+\frac{1}{2}} - H_{z-\frac{1}{2}}^{t+\frac{1}{2}}}{\Delta z} + P_z^{t+1} - P_{z}^{t},
\end{eqnarray}
where $E^i_j=E_x(j\Delta z,i\Delta t)$ and $\frac{1}{2}$ subscripts indicate that the electric and magnetic field numerical grids are offset by half a time step and the evolution equations are interleaved in time. This guarantees second-order accuracy of the scheme \cite{Yee}.

The optical properties of the medium result from the dynamics of the polarization $P$. The evolution of the polarization is calculated with the use of the so-called Auxiliary Differential Equations (ADE) method \cite{Alsunaidi}. The polarization is described by a differential equation of a harmonic oscillator
\begin{equation}\label{polaryzacje}
\ddot{P}+\gamma\dot{P}+\omega^2_{0} P=\frac{f}{\epsilon_b} E
\end{equation}
with parameters $\epsilon_b$, $f$, $\gamma$ characterizing the optical medium. In particular, for the excitonic resonance, $\hbar\omega_0$ is the exciton energy, $\gamma$ is the spectral width of the exciton absorption line and $f$ is the so-called oscillator strength calculated from the transition dipole moment. In the frequency domain, the above model corresponds to
\begin{eqnarray}\label{drude}
\vec{P}(\omega)&=&\epsilon(\omega)\vec{E}(\omega),\nonumber\\
\epsilon(\omega)&=&\epsilon_b+\sum_{j=1}^{2}\frac{f_j}{\omega_{0j}^2 - \omega^2 - i\gamma_j\omega}
\end{eqnarray} 
where two oscillators $j=1,2$ are used to represent the two exciton resonances. Crucially, the medium parameters are not static; the exciton energy $\hbar\omega_0$ is subject to a shift by $\Delta E_{VdW}$ which, in turn, depends on the dynamic populations $\sigma_{ij}$.

To sum up, the calculation procedure is as follows:
\begin{itemize}
\item Given current values of the density matrix elements $\sigma_{ij}^{(t)}$ and the field $\varepsilon^{(t)}$, calculate new elements $\sigma_{ij}^{(t+\Delta t)}$ with Eqs. (\ref{sigma_num}).
\item Use the populations to estimate the Rydberg blockade energy $\Delta E_{VdW}$.
\item With discretized version of Eq. (\ref{drude}), calculate medium polarization.
\item Calculate new electric and magnetic field values with Eqs. (\ref{FDTD}). 
\end{itemize}

\begin{figure}[ht!]
\includegraphics[width=.95\linewidth]{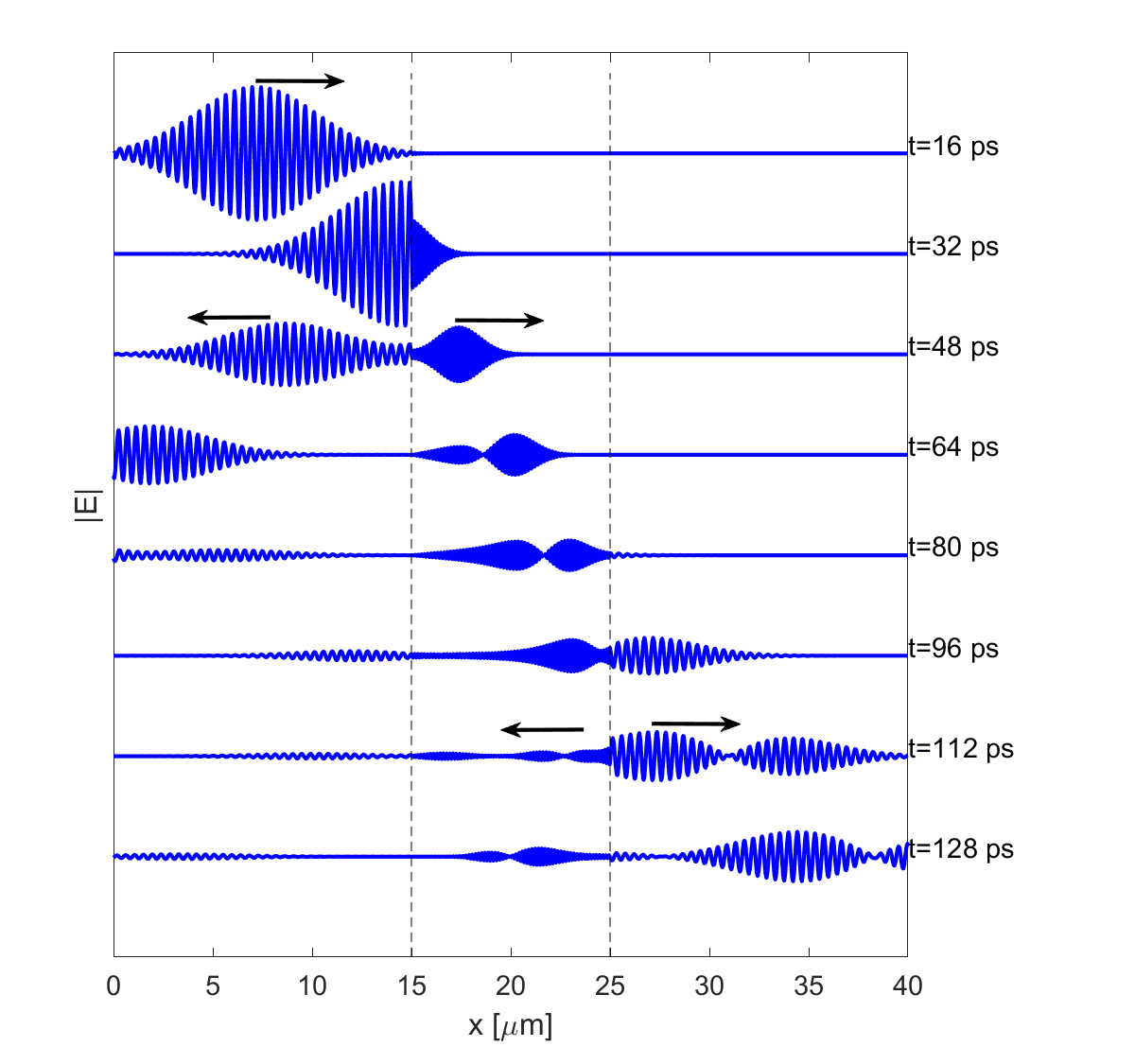}
\caption{Snapshots of the electric field distribution obtained in the simulation after indicated time. Vertical, dashed lines represent the boundaries of the Cu$_2$O crystal.}\label{fig:snapshots}
\end{figure}

The Fig. \ref{fig:snapshots} represents a typical calculation result, e.g. the time evolution of the field propagating through the system.

On the top panel (t=16 fs), one can See a Gaussian pulse propagating to the right. Next, at t=32 fs, it has already partially entered the crystal. At t=48 fs, one can distinguish the transmitted and reflected parts of the pulse, propagating to the right and left, correspondingly. At t=64 fs, one can see that the pulse has split into two distinct parts. The faster mode that has propagated further into the crystal represents the center of the impulse's spectrum, tuned exactly to the excitonic resonance and propagating in the conditions of anomalous dispersion. Initially, this fast mode contains the majority of the incident energy and has larger amplitude (t=64 fs). However, at t=80 fs, the amplitudes of the fast and slow modes are already approximately equal. At t=96 ps, the fast mode is exiting the crystal, with the slow mode exiting at t=112 fs. Finally, at t=128 fs, one can see the part of the impulse reflected from the inner wall of the crystal.

\end{document}